\documentstyle[preprint,aps]{revtex}

\tighten
\begin{document}

\title{
       Properties of $^{12}$C in the {\it ab initio} nuclear shell-model
}
\medskip

\author{
	P. Navr\'atil$^{a,b}$,
         J. P. Vary$^{c}$
         and B. R. Barrett$^{a}$
        }

\medskip

\address{
	$^a$Department of Physics, University of Arizona,
        	Tucson, Arizona 85721\\
           $^b$Nuclear Physics Institute, 
	       AS CR, 
               	250 68 \v{R}e\v{z} near Prague, Czech Republic \\
           $^c$Department of Physics and Astronomy, and\\
       International Institute of Theoretical and Applied Physics,\\
	        Iowa State University, Ames, Iowa 50011 
}

\maketitle

\bigskip

\begin{abstract}
We obtain properties of $^{12}$C in the {\it ab initio} no-core
nuclear shell-model. The effective Hamiltonians are derived microscopically
from the realistic CD-Bonn
and the Argonne V8' nucleon-nucleon (NN) potentials as a function
of the finite harmonic oscillator basis space. Binding energies,
excitation spectra and electromagnetic properties 
are presented for model spaces up to $5\hbar\Omega$.  
The favorable comparison with available data is a
consequence of the underlying NN interaction rather than a
phenomenological fit.
\end{abstract}

\bigskip
\bigskip
\bigskip

\narrowtext



While various methods have been developed to solve the
three- and four-nucleon systems with
realistic interactions \cite{Fad60,Ya67,VKR95,GFMC},
few approaches are suitable for heavier nuclei
at this time.
Apart from the coupled cluster method \cite{CCM} applied 
to closed-shell and near-closed shell nuclei,
the Green's function Monte Carlo method is the only approach
for which exact solutions of systems with $A\le8$,
interacting by realistic potentials, have been obtained \cite{GFMC}.

For more complex nuclei, treated as systems of nucleons
interacting by realistic NN interactions,
we apply the no-core shell-model approach \cite{ZBV,NB96,NB98,NKB99}.
To date, this {\it ab initio} method has been successfully
applied to solve the three-nucleon as well as the four-nucleon
bound-state problem \cite{NB98,NKB99}. Here, we address
a vastly more
complex system, $^{12}$C, and present first results for an
illustrative set of observables with two realistic NN 
interactions.

There are several pressing reasons to investigate $^{12}$C
in a way that preserves as much predictive power as possible.
The $^{12}$C nucleus plays an important role \cite{HeyesVogel}
in neutrino studies using liquid scintillator detectors.
Also, there has been considerable interest recently in parity-violating
electron scattering from $(J^\pi, T)=(0^+, 0)$
targets, like $^{12}$C, 
to measure the strangeness content of the nucleon \cite{MD92,EO99}.
For these and many other reasons,
there have been multi-$\hbar\Omega$
shell model studies of $^{12}$C
in the past \cite{Kara95,SS98,HT99}.  However, unlike our approach,
phenomenological effective interactions were used.

We start from the
two-body Hamiltonian for the $A$-nucleon system, which depends
on the intrinsic coordinates alone,
$H_A=T_{rel} + {\cal V}$, where
$T_{rel}$ is the relative kinetic energy
operator and ${\cal V}$ is the sum of two-body nuclear and Coulomb
interactions, ${\cal V} = V_{\rm N} + V_{\rm C}$.  There is no
phenomenological one-body term. We neglect
many-body interactions at present.  To facilitate
our work, we add an $A$-nucleon
Harmonic Oscillator (HO) Hamiltonian acting solely on the
center-of-mass (CM), $H_{\rm CM}=T_{\rm CM}+U_{\rm CM}$,
where $U_{\rm CM}=\frac{1}{2}Am\Omega^2 \vec{R}^2$,
$\vec{R}=\frac{1}{A}\sum_{i=1}^{A}\vec{r}_i$
and $m$ is the nucleon mass.
The effect of this HO CM Hamiltonian will be 
subtracted in the final many-body calculation.
The Hamiltonian, with a pseudo-dependence on
$\Omega$, can be cast into the form
%
\begin{eqnarray}\label{hamomega}
H_A^\Omega &=& \sum_{i=1}^{A} h_i + \sum_{i<j=1}^{A} V_{ij}
=
\sum_{i=1}^A \left[ \frac{\vec{p}_i^2}{2m}
+\frac{1}{2}m\Omega^2 \vec{r}^2_i
\right]
\nonumber \\
&&
+ \sum_{i<j=1}^A \left[ {\cal V}_{ij}
-\frac{m\Omega^2}{2A}
(\vec{r}_i-\vec{r}_j)^2
\right] \; .
\end{eqnarray}
%
Since we solve the many-body problem in a finite HO model space,
the realistic nuclear interaction in 
(\ref{hamomega}) will
yield pathological results unless we
derive a model-space dependent effective Hamiltonian.
For this purpose, we adopt approaches presented
by Lee and Suzuki \cite{LS80}, Da Providencia and Shakin \cite{PS64},
and Suzuki and Okamoto \cite{UMOA},
which yield an Hermitian effective Hamiltonian.

According to Da Providencia and Shakin \cite{PS64}, a
unitary transformation
of the Hamiltonian $H_A^\Omega$, which is able to accommodate
the short-range two-body
correlations,
can be introduced by choosing a two-body, 
in our case translationally invariant, 
antihermitian operator $S=\sum_{i<j=1}^A S_{ij}$,
such that
\begin{equation}\label{UMOAtrans}
{\cal H} = e^{-S} H_A^\Omega e^{S} \; .
\end{equation}
The transformed Hamiltonian can be 
expanded in terms of up to $A$-body
clusters
${\cal H} = {\cal H}^{(1)} + {\cal H}^{(2)} + {\cal H}^{(3)} +\ldots \; ,$
where the one-body and two-body pieces are given as
${\cal H}^{(1)} = \sum_{i=1}^{A} h_i \; ,$
${\cal H}^{(2)} = \sum_{i<j=1}^{A} \tilde{V}_{ij} \; ,$
with
\begin{equation}\label{UMOAexplterms}
\tilde{V}_{12} = e^{-S_{12}}(h_1+h_2+V_{12})e^{S_{12}}-(h_1+h_2)
\; .
\end{equation}
The full space is divided into a model or P-space, and a Q-space, using
the projectors $P$ and $Q$ with $P+Q=1$.
It is then possible to determine the transformation operator $S_{12}$
from the decoupling condition
\begin{equation}\label{UMOAdecoupl}
Q_{2} e^{-S_{12}}(h_1+h_2+V_{12})e^{S_{12}} P_{2} = 0 \; .
\end{equation}
The two-nucleon-state projectors ($P_2, Q_2$)
follow from the
definitions of the $A$-nucleon projectors $P$, $Q$.
This approach has a solution \cite{UMOA},
$S_{12} = {\rm arctanh}(\omega-\omega^\dagger)    \; ,$
with the operator $\omega$ satisfying $\omega=Q_2\omega P_2$.
This is the same operator, which we previously employed
\cite{NB96,NB98,NKB99}. It can be directly obtained
from the eigensolutions $|k\rangle $ of $h_1+h_2+V_{12}$ 
as $\langle\alpha_Q|\omega|\alpha_P\rangle = \sum_{k \in{\cal K}}
\langle\alpha_Q|k\rangle\langle\tilde{k}|\alpha_P\rangle $,
where we denote by tilde the inverted matrix of 
$\langle\alpha_P|k\rangle$. 
In the above relation, $|\alpha_P\rangle$ and $|\alpha_Q\rangle$
are the 2-particle model-space and Q-space 
basis states, respectively,  
and ${\cal K}$ denotes a set of $d_P$ eigenstates, whose properties
are reproduced in the model space, 
with $d_P$ equal to the dimension of the 
model space.

The resulting two-body effective interaction $\tilde{V}_{12}$
depends on $A$, on the HO frequency $\Omega$ and on $N_{\rm max}$,
the maximum many-body HO excitation energy (above the
lowest configuration) defining the P-space.
It follows that ${\cal H}^{(1)} + {\cal H}^{(2)} - H_{\rm CM}$ 
is translationally invariant and that
$\tilde{V}_{12}\rightarrow V_{12}$
for $N_{\rm max}\rightarrow \infty$.
A significant consequence of preserving translational invariance is 
the factorization of our wave function into a product of a CM 
$\frac{3}{2}\hbar\Omega$ 
component times an internal component.
Hence, it is straightforward to correct
exactly any observable for the CM effects. This
feature distinguishes our approach from most phenomenological shell
model studies that involve multiple HO shells. 

The most significant approximation used in the present
application is the neglect
of higher than two-body clusters in the unitary transformed Hamiltonian
expansion.
Our method is not a variational approach so the neglected
clusters can contribute either positively or negatively to
the binding energy. Indeed, we find that the
character of the convergence depends
on the choice of $\Omega$ \cite{ZBV,NB98,NKB99}.
The method can be readily generalized in order to include, e.g.,
three-body clusters, and to demand the model-space decoupling on
the three-body cluster level. Such a generalization leads to the
derivation of the three-body effective interaction, which has been
successfully
applied in our calculations for the $A=4$ system \cite{NB98,NKB99}.
We learned from these studies that
the contribution of higher order clusters to the ground-state energy
are about 10\% in similar model spaces that we employ here, when an optimal
HO frequency is chosen, as it is the case in the present application.

To solve for the properties of $^{12}$C,
we employ the m-scheme Many-Fermion Dynamics code \cite{VZ94}.
Due to the fast growing matrix dimensions,
reaching 6 488 004 at the $N_{\rm max}=5$ model space,
we are restricted to $N_{\rm max}=0, 2, 4$
for the positive-parity states and $N_{\rm max}=1, 3, 5$ for
the negative-parity states. Here, we utilize $\hbar\Omega=15$ MeV which
lies in the range where the
largest model space results are least sensitive to
$\hbar\Omega$.  Full details will be reported elsewhere.

We present results for the CD-Bonn \cite{Machl}
and the Argonne V8' \cite{GFMC} NN potentials.
Our positive-parity state results are presented in Table \ref{tab1}
and in Fig. \ref{figc12hw420}, and the negative-parity state results
are in Table \ref{tab2} and Fig. \ref{figc12hw531}.
While the energy of the lowest eigenstate of each parity increases with
increasing model space, the relative level spacings are
less dependent on model space size. 
As a gauge of trends with increasing model space
size, consider the rms changes
in excitation energies of the first 7 excited states
of each parity in the CD-Bonn case.
For positive parity states, the rms changes are
1.31 (0.22) MeV in going from 0 to 2 (2 to 4)$\hbar\Omega$. 
For negative parity states, the rms 
changes are 0.87 (0.20) MeV in going from 1 to 3 (3 to 5)$\hbar\Omega$. 
The difference between the $N_{\rm max}=2(3)$
and 4(5) results is significantly
smaller than that between the $N_{\rm max}=0(1)$
and 2(3) results which is similar to the convergence 
trends we saw in lighter systems \cite{ZBV,NB98,NKB99}.
Our obtained binding energy of about 88 MeV in the $4\hbar\Omega$
space is expected to decrease with a further model space enlargement.
We estimate, however, that our result should be within 10\% of
the exact solution for the two-body NN potential used. In order
to reach the experimental binding energy, likely a true three-body
NN interaction is neccessary \cite{GFMC}.

In general, we obtain a reasonable agreement of the
states dominated by $0\hbar\Omega$
and $1\hbar\Omega$ configurations with experimental levels.
We also observe a general trend of improvement with increasing
model space size, i.e., the ordering of the $T=1$ states. We obtain a 
resonable set of excitation energies for the $T=1$ states relative
to the lowest $T=0$ state of each parity. In addition,
our lowest $0^+$ $T=2$ state lies between 27 and 29 MeV,
depending on the NN potential and the model space,
in good agreement with the experimental $0^+ 2$ state at 27.595 MeV. 
We note that the favorable comparison with available data is a
consequence of the underlying NN interaction rather than a
phenomenological fit.
Our ground-state wave function in the $4\hbar\Omega$ calculation
contains about 61\% of the $0\hbar\Omega$ component. The occupancy of
the $0p3/2$ level is about 5.74 nucleons, while the 
occupancy of the $0p1/2$ level is about 1.90 nucleons.
From Tables \ref{tab1} and \ref{tab2},
it is clear that the excitation energies of the negative-parity
states relative to the positive-parity states decrease rapidly
with the model-space enlargement. 
Still, even in our largest spaces
the $3^- 0$ state is more than 5 MeV too high compared to the experiment.

In order to achieve a more realistic
excitation energy a still larger HO expansion is needed especially
for states with significant cluster structure.
The two- and higher-$\hbar\Omega$ dominated
states, such as the 7.65 MeV $0^+ 0$ state
that is known to be a three-alpha cluster resonance \cite{POCM}, 
are not seen in the low-lying part of our calculated spectra.
In general, the convergence rate of the
$2\hbar\Omega$ dominated states is quite different than that of the 
ground state as we observed in $^4$He calculations performed in the present
formalism \cite{NB98,NKB99}. Also, an optimal HO frequency for the convergence 
of the ground
state will differ from the optimal frequency for the $2\hbar\Omega$ states.
We investigated the position of the lowest $2\hbar\Omega$ dominated states
and the giant-quadrupole resonance (GQR) E2 distribution.
Our lowest $2\hbar\Omega$ $0^+$ state lies at about 40 MeV excitation energy
and the GQR E2 strength is fragmented between 43 to 50 MeV
in the $2\hbar\Omega$ calculation. In the $4\hbar\Omega$ model space
the excitation energy of the lowest $2\hbar\Omega$ $0^+$ state drops by 5 MeV
to about 35 MeV and similarly the GQR strength position is lowered to 37-47 MeV.
We note that the experimental GQR strength is observed in the range 18-28 MeV 
\cite{DCW}.

There is little difference between the results
from the two NN interactions, although the overall
agreement with experiment is slightly better for the CD-Bonn
NN potential, in particular for the $T=1$ states.

Our radius and E2 results, based on the
bare radius operator and bare nucleon charges, are
smaller than the experimental values.
The underestimation of the rms radius, the quadrupole moment and
the E2 transitions is linked with the overestimation of the position 
of the GQR strength and suggests that even in the $N_{\rm max}=4$ 
space we still miss significant clustering effects.
We also observe a strong model space dependence of the M1 transitions,
$1^+1 \rightarrow 0^+ 0$.
Clearly, there is still a need for
effective operators, which are calculable
within our theoretical framework.
In general, to compute a two-body correction to a one-body operator
in our formalism is more involved than the evaluation of the
effective interaction.
But, it is easy to study the lowest order renormalization for
a two-body operator depending on the relative position of two nucleons
as , e.g., the point-nucleon rms radius operator.
Then, $O_{\rm eff}\approx
\sum_{i<j=1}^A  e^{-S_{ij}} O_{ij} e^{S_{ij}}$. We computed this 
term for the point-proton rms radius operator and found that
the renormalization leads to an increase of the radius and that the size 
of this increase drops as the model space size increases. 
The $r_p$ results
presented in Table \ref{tab1} that were obtained without renormalization
should be increased due to the renormalization by about 0.06, 0.02
and 0.01 fm for the $N_{\rm max}=0, 2$ and 4
model spaces, respectively. This does not imply that the renormalization 
of other operators, e.g., the E2 operator, cannot be substantially higher. 
Similarly, as observed in our 
$^3$H calculations
\cite{NKB99}, 
we anticipate that, in contrast with the 
energies,
the higher-order corrections
will be more significant and the overall convergence slower for 
other observables.

We present these results as a useful description of
the 0 and $1\hbar\Omega$-dominated states of $^{12}$C.
Our wave functions along with the one-body and two-body
densities may also be used to predict cross sections
for neutrino and muon reactions with $^{12}$C.
Such cross sections will be the subject of future investigations.
The trends are encouraging and we will carry out larger
model space investigations in order to achieve greater
convergence.

\acknowledgements{
This work was supported in part by the NSF grant No. PHY96-05192
and in part by USDOE grant No. DE-FG-02-87ER-40371,
Division of High Energy and Nuclear Physics. JPV acknowledges
valuable technical assistance from J.J. Coyle and D.R. Fils.
}

\begin{figure}
\caption{Experimental and theoretical positive-parity excitation spectra of
$^{12}$C. Results obtained in $4\hbar\Omega$, $2\hbar\Omega$
and $0\hbar\Omega$
model spaces are compared. The effective interaction was derived
from the CD-Bonn NN potential in a HO basis with
$\hbar\Omega=15$ MeV. The experimental values are 
from Ref. \protect\cite{AS90}.
}
\label{figc12hw420}
\end{figure}

\begin{figure}
\caption{Experimental and theoretical negative-parity 
spectra of
$^{12}$C. Results obtained in $5\hbar\Omega$, $3\hbar\Omega$
and $1\hbar\Omega$ model spaces are compared.
Other factors are the same as in Fig. \protect\ref{figc12hw420}.
}
\label{figc12hw531}
\end{figure}

\newpage
\widetext

\begin{table}
\begin{tabular}{cccccccc}
 & $^{12}$C & & CD-Bonn &  &  & AV8' &  \\ 
 model space & - & $4\hbar\Omega$ & $2\hbar\Omega$ & $0\hbar\Omega$ & $4\hbar\Omega$ &
$2\hbar\Omega$ & $0\hbar\Omega$ \\
\hline
$|E_{\rm gs}|$ [MeV]   & 92.162 & 88.518 & 92.353 & 104.947  & 87.675 & 92.195 & 104.753  \\
$r_p$ [fm]             & 2.35(2)& 2.199  & 2.228  & 2.376    & 2.202  & 2.228  & 2.376  \\
$Q_{2^+}$ [$e$ fm$^2$] & +6(3)  & 4.533  & 4.430  & 4.253    & 4.536  & 4.427  & 4.250  \\
\hline
$E_{\rm x}(0^+ 0)$ [MeV] & 0.0      & 0.0    & 0.0     & 0.0    & 0.0    & 0.0    & 0.0     \\
$E_{\rm x}(2^+ 0)$ [MeV] & 4.439    & 3.697  & 3.837   & 3.734  & 3.584  & 3.766  & 3.699   \\
$E_{\rm x}(1^+ 0)$ [MeV] & 12.710   & 14.141 & 14.525  & 13.866 & 14.044 & 14.549 & 13.935  \\
$E_{\rm x}(4^+ 0)$ [MeV] & 14.083   & 13.355 & 13.636  & 12.406 & 12.848 & 13.255 & 12.192  \\
$E_{\rm x}(1^+ 1)$ [MeV] & 15.110   & 16.165 & 16.291  & 15.290 & 16.295 & 16.515 & 15.488  \\
$E_{\rm x}(2^+ 1)$ [MeV] & 16.106   & 17.717 & 17.945  & 15.970 & 17.945 & 17.823 & 15.920  \\
$E_{\rm x}(0^+ 1)$ [MeV] & 17.760   & 16.618 & 16.493  & 14.698 & 16.205 & 16.208 & 14.574  \\
\hline
B(E2;$2^+0 \rightarrow 0^+0$) & 7.59(42)   & 4.625  & 4.412  & 4.092  & 4.612  & 4.397  & 4.091 \\
B(M1;$1^+0 \rightarrow 0^+0$) & 0.0145(21) & 0.0042 & 0.0032 & 0.0013 & 0.0026 & 0.0019 & 0.0008  \\
B(M1;$1^+0 \rightarrow 2^+0$) & 0.0081(14) & 0.0017 & 0.0013 & 0.0008 & 0.0013 & 0.0012 & 0.0008 \\
B(M1;$1^+1 \rightarrow 0^+0$) & 0.951(20)  & 0.355  & 0.280  & 0.158  & 0.316  & 0.252  & 0.147 \\
B(M1;$1^+1 \rightarrow 2^+0$) & 0.068(9)   & 0.0002 & 0.0028 & 0.0115 & 0.0023 & 0.0078 & 0.0167 \\
B(E2;$2^+1 \rightarrow 0^+0$) & 0.65(13)   & 0.283  & 0.015  & 0.0018 & 0.104  & 0.000  & 0.002
\end{tabular}
\caption{Experimental and calculated binding energies, ground-state
point-proton rms radii, the $2^+_1$-state quadrupole moments,
as well as E2 transitions, in $e^2$ fm$^4$, and M1 transitions, in $\mu_N^2$,
of $^{12}$C. Results obtained in different model spaces, i.e.,
$N_{\rm max}=4, 2, 0$,
and using effective interactions derived from
the CD-Bonn and the Argonne V8' NN potentials are compared.
A HO frequency $\hbar\Omega=15$ MeV was employed.
The experimental values are 
from Refs. \protect\cite{AS90,T88}.}
\label{tab1}
\end{table}

\begin{table}
\begin{tabular}{cccccccc}
 & $^{12}$C & & CD-Bonn  &                &    & AV8'    &   \\
 model space              & -        & $5\hbar\Omega$ & $3\hbar\Omega$
&$1\hbar\Omega$   & $5\hbar\Omega$ &
$3\hbar\Omega$ & $1\hbar\Omega$ \\
\hline
$|E(3^- 0)|$ [MeV]  & 82.521 & 72.952 & 75.331 & 83.390 & 72.300 & 75.360 & 83.459   \\
$r_p$ [fm]          &        & 2.309  & 2.316  & 2.425  & 2.310  & 2.315  & 2.425    \\
$Q_{3^-}$ [$e$ fm$^2$] &     & -7.942 & -7.596 & -6.936 & -7.920 & -7.575 & -6.933   \\
$E(3^- 0)-E_{\rm gs}$ [MeV]  & 9.641  & 15.566 & 17.022 & 21.557 & 15.375 & 16.835 & 21.294 \\
\hline
$E_{\rm x}(3^- 0)$ [MeV] & 0.0  & 0.0 & 0.0    & 0.0    & 0.0    & 0.0    & 0.0       \\
$E_{\rm x}(1^- 0)$ [MeV] & 1.203 & 2.093 & 2.256 & 1.561 & 2.112 & 2.274  & 1.552     \\
$E_{\rm x}(2^- 0)$ [MeV] & 2.187 & 3.722 & 4.051 & 3.582 & 3.722 & 4.057  & 3.567     \\
$E_{\rm x}(4^- 0)$ [MeV] & 3.711 & 4.866 & 5.084 & 4.768 & 4.741 & 4.993  & 4.710     \\
$E_{\rm x}(0^- 0)$ [MeV] &       & 7.148 & 7.062 & 5.712 & 7.148 & 7.156  & 5.777     \\
$E_{\rm x}(2^- 1)$ [MeV] & 6.929 & 7.671 & 7.783 & 7.340 & 7.949 & 8.237  & 7.574    \\
$E_{\rm x}(3^- 0)$ [MeV] &       & 7.877 & 8.151 & 6.886 & 7.651 & 7.983  & 6.745     \\
$E_{\rm x}(1^- 1)$ [MeV] & 7.589 & 8.048 & 7.951 & 7.042 & 8.117 & 8.096  & 7.184
\end{tabular}
\caption{Experimental and calculated negative-parity state energies,
the $3^- 0$-state point-proton rms radii, 
and quadrupole moments are shown.
Results obtained in different model spaces, i.e.,
$N_{\rm max}=5, 3, 1$,
and using effective interactions derived from
the CD-Bonn and the Argonne V8' NN potentials are compared.
The calculated excitation energy of $3^- 0$ is obtained by comparing its
energy in the $N\hbar\Omega$ space with the ground state 
in the $(N-1)\hbar\Omega$ space.
A HO frequency $\hbar\Omega=15$ MeV was employed.
The experimental values are taken from Ref. \protect\cite{AS90}.}
\label{tab2}
\end{table}

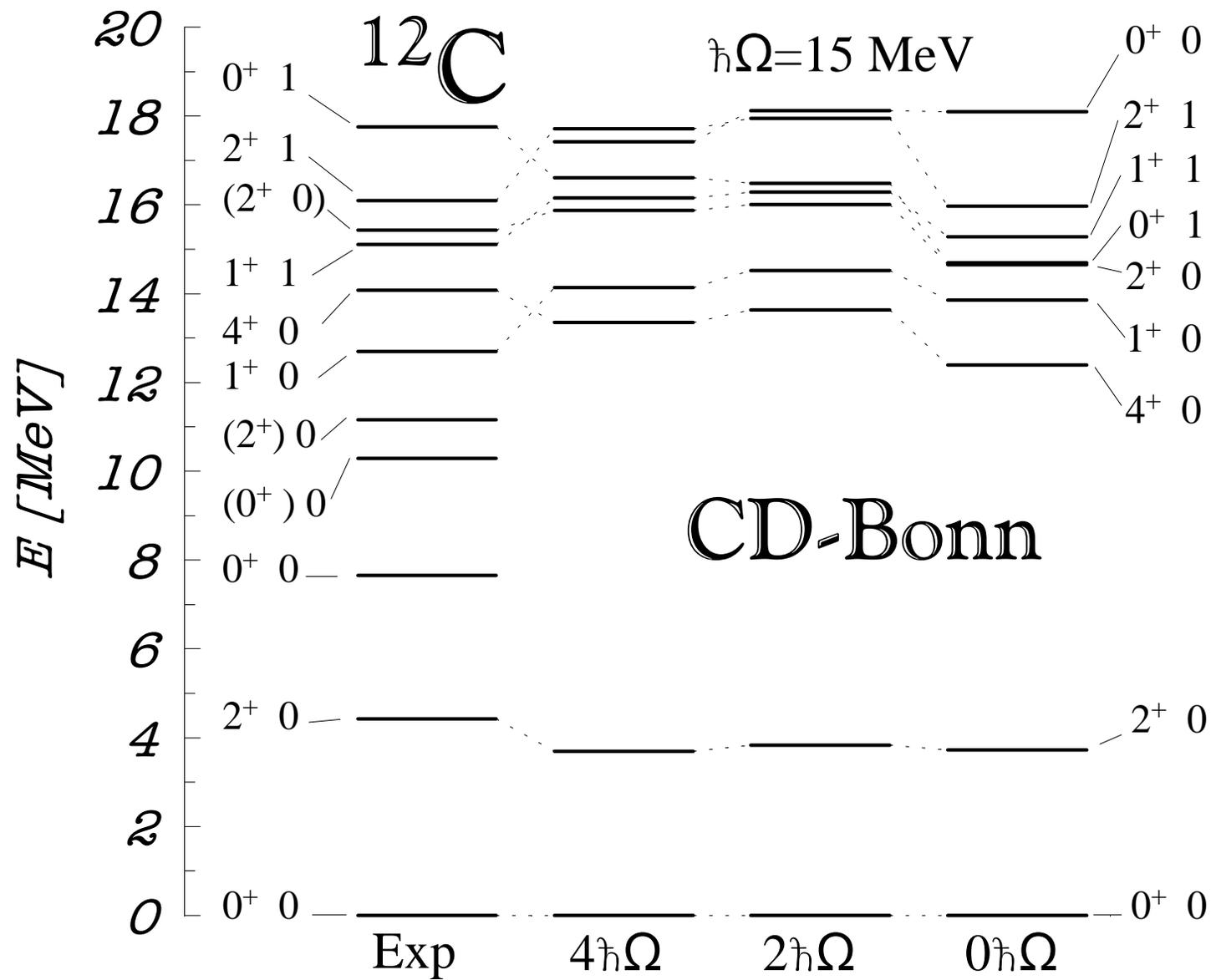
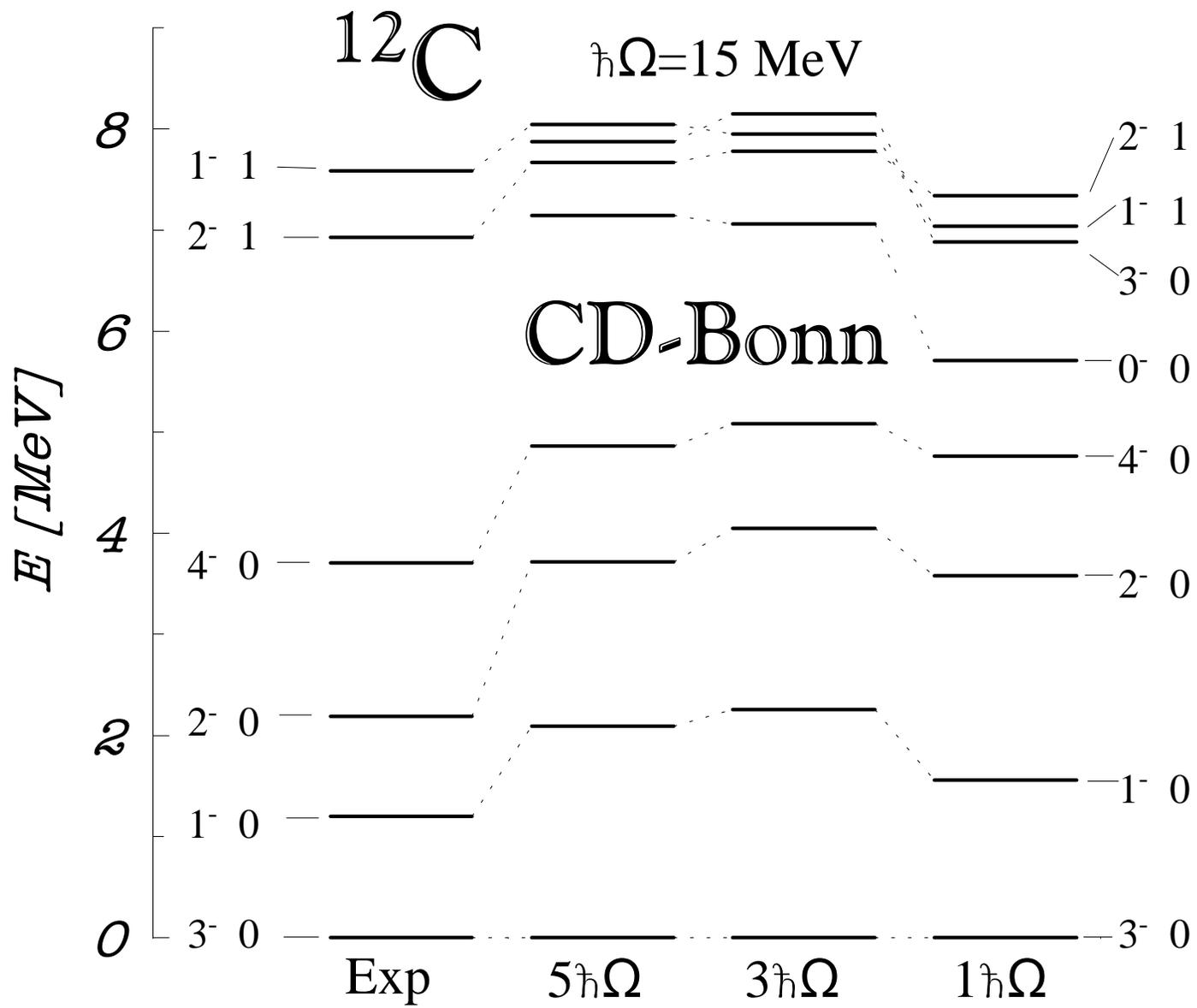
\end{document}